\documentstyle[psfig,twocolumn,aps,graphicx,prl,rotating,epsfig,amsfonts,enumerate]{revtex}

\tighten
\begin{document}
\setlength{\unitlength}{1mm}

\draft

\title{ SrCu$_2$(BO$_3$)$_2$ -- a Two Dimensional Spin Liquid} 

\author{Christian Knetter, Alexander B\"uhler,
  Erwin M\"uller-Hartmann and G\"otz S. Uhrig}

\address{Institut f\"ur Theoretische Physik, Universit\"at zu
  K\"oln, Z\"ulpicher Str. 77, D-50937 K\"oln, Germany\\[1mm]
  {\rm(\today)} }

\maketitle

\begin{abstract}
We study an extended Shastry-Sutherland model for
SrCu$_2$(BO$_3$)$_2$ and analyze the low lying parts of the energy spectrum by
means of a perturbative unitary transformation based on flow
equations. The derivation of the 1-magnon dispersion (elementary
triplets) is discussed. Additionally, we give a
quantitative description (symmetries and energies) of bound states
made from two elementary triplets. Our high order results allow to fix
the model parameters for SrCu$_2$(BO$_3$)$_2$ precisely:  $J_1=6.16(10)$meV, 
$x:=J_2/J_1=0.603(3)$, $J_\perp=1.3(2)$meV. To our knowledge this is
the first quantitative treatment of bound states in a true 2d model.
\end{abstract}
\begin{figure}[ht]
\begin{center}
{\psfig{figure=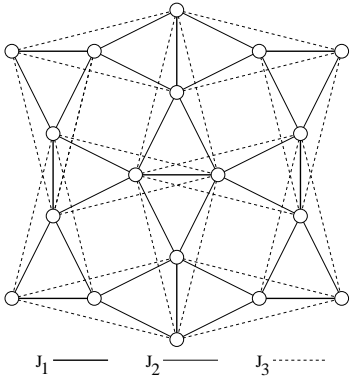 ,height=50mm}}
\end{center}
\caption{\footnotesize{A part of the extended Shastry-Sutherland
    model. The circles are $S=1/2$ entities. For 
    $J_3=0$ this model reduces to the original Shastry-Sutherland model. The
    coupling $J_1$ is assumed to be antiferromagnetic. The starting
    point of our analysis is the limit of strong dimerization
    ($J_2,J_3 \rightarrow 0$).} \label{model}} 
\end{figure}
\begin{figure}[ht]
\begin{center}
{\psfig{figure=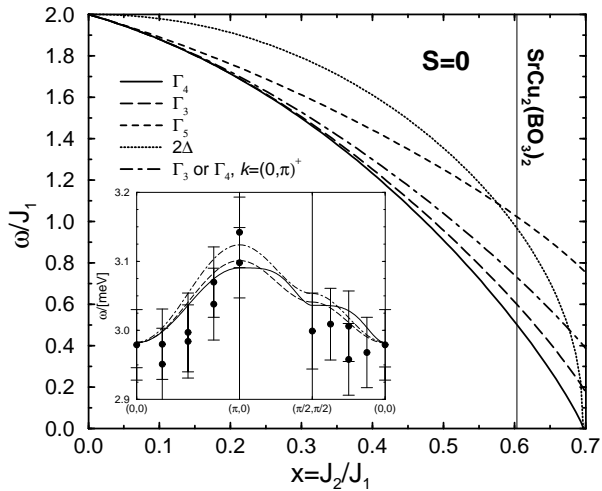 ,height=65mm}}
\end{center}
\vspace*{-1mm}
\caption{\footnotesize{Energy of the lowest-lying $S=0$ states. Curves
    refer to ${\mathbf k}={\mathbf 0}$ except the dashed-dotted
    one. The dotted curve displays the continuum at $2\Delta$. Inset:
    1-triplet dispersion for various values of ($J_1$/meV;$x$;$y$):
    (6.56;0.615;0) dashed-dotted line, (6.67;0.59;-0.05) dashed line, (6.16;0.603;0)
    solid line.} \label{merge1}}
\end{figure}
\begin{figure}[ht]
\begin{center}
{\psfig{figure=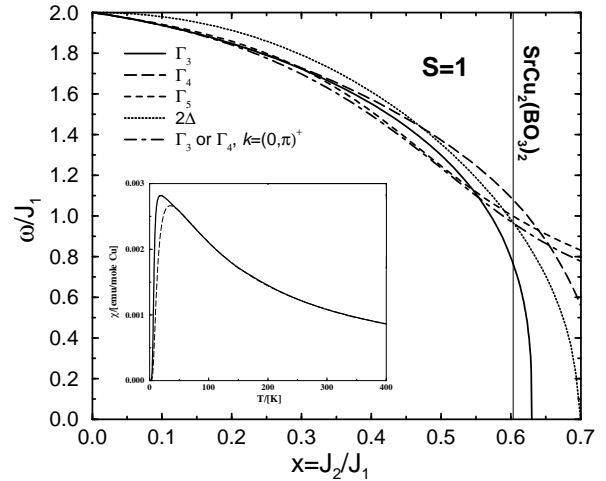 ,height=65mm}}
\end{center}
\vspace*{-1mm}
\caption{\footnotesize{As in Fig.~\protect\ref{merge1} for  $S=1$.
      Inset: Magnetic susceptibility. Theory (dashed)
      with directional rms average $g=2.13$ \protect\cite{nojir99},
      $x,J_1$ as in Fig.~\protect\ref{merge1};
      experiment (solid) on powder \protect\cite{kagey99}.} \label{merge2}}
\end{figure}

\pacs{PACS numbers: 75.40.Gb, 75.30.Kz, 75.50.Ee, 75.10.Jm}
\narrowtext
Presently low dimensional quantum antiferromagnets are investigated
intensively both in experiment and theory. Systems
that do not show a long ranged ordered ground state, so called spin
liquids,  are particularly interesting. Besides low spin and
low coordination number spin liquids are more likely to form in
strongly frustrated geometries.
The recently synthesized antiferromagnetic material
SrCu$_2$(BO$_3$)$_2$~\cite{smith91,kagey99} is a nice two dimensional
example since it is a realization of the Shastry-Sutherland
model~\cite{miyah98,shast81b} (cf. Fig.~\ref{model}).
In this model frustration plays an essential role. Each spin is
coupled to pairs of spins on dimers. If these pairs form singlets
the couplings between dimers is without effect and the
singlet-on-dimers product state is always an eigen state and for
certain parameters the ground state
\cite{broek80,shast81a,shast81b,mulle00}. In this {\it dimer phase} the system is gaped. The
Hamiltonian reads  
\begin{equation}
 \label{model_ham}
{H}=J_1\!\!\underbrace{\sum_{<i,j>}\! \vec S_i\cdot \vec S_j}_{H_0}
+J_2\!\!\underbrace{\sum_{<i,k>} \!\vec S_i\cdot \vec S_k}_{H_1}
+J_3\!\!\underbrace{\sum_{<i,l>} \!\vec S_i\cdot \vec S_l}_{H_2},
\end{equation}
where the bonds corresponding to
interactions $J_1$, $J_2$ and $J_3$ are shown in Fig.~\ref{model}.\\
For larger ratios $x=J_2/J_1$ and $y=J_3/J_1$ the system transits into
a N\'eel phase, competing with a helical phase for certain values of
$y$. Finally there exists a ferromagnetic phase in the region of
larger negative $x$ and $y$. A profound analysis of these scenarios can
be found in Ref.~\cite{mulle00}. Along the $x$-axis Koga and
Kawakami~\cite{koga00} found an additional plaquette-singlet 
phase, which we see as beeing related to the helical phase.\\
In this article we start from the dimer phase where an elementary
excitation is given by breaking up $one$ singlet and substituting a
triplet instead, which acquires a dispersion by hopping from dimer to
dimer. For the next higher excitations we focus on bound states formed
from pairs of the elementary triplets. We will perturbatively
calculate the low lying excitations of the model about the limit of
strong dimerization. The problem can be
stated as
\begin{equation}
  \label{H_pert}
  \frac{H}{J_1} = H_0 + x H_{\rm S}\ ,\ \ \mbox{with}\ H_{\rm S} = H_1
  + \frac{y}{x} H_2\ .
\end{equation}
In the limit of isolated dimers ($x=0$, with $y/x$
finite) $H$ is bounded from below and has an equidistant energy
spectrum, since $H_0$ simply counts the number of
excited dimers (up to a trivial constant). Furthermore $H_{\rm S}$ can be
decomposed
\begin{equation}
  \label{H_S_decomp_1}
  H_{\rm S}=T_{-1}+T_{0}+T_1\ ,
\end{equation}
where $T_i$ creates $i$ (destroys for $i<0$) elementary
triplets (energy quanta). These properties of $H_0$ and $H_{\rm S}$
allow us to use the perturbative
unitary transformation \cite{knett99a} based on flow equations
\cite{wegne94}. This technique enables us to link smoothly and uniquely $H$ at
$x \neq 0$ to an effective  $H_{\rm eff}$ conserving the
number of  triplets on dimers $[H_{\rm eff},H_0]=0$. This permits
a clear distinction between the ground state (without triplets),
 the 1-triplet sector, the 2-triplet sector etc.. Thus the effective
Hamiltonian is block diagonal. The explicit form is given by
\begin{equation}
\label{H_eff}
{\cal H}_{\rm eff} = {\cal H}_0 +\sum_{k=1}^{\infty}x^{k} 
\sum_{|\underline{m}|=k, M(\underline{m})=0} C(\underline{m}) 
T(\underline{m})\ ,
\end{equation}
where $\underline{m}$ is a vector of dimension $k$ of which the
components are in $\{\pm 1, 0\}$; $M(\underline{m})=0$
signifies 
that the sum of the components vanishes which reflects the conservation of the
number of energy quanta. The operator product
$T(\underline{m})$ is defined by
$T_{\underline{m}}=T_{m_1}T_{m_2}\cdots T_{m_k}$. The coefficients 
$C(\underline{m})$ are generally valid fractions, which we computed up
to order $k=15$.\\
In terms of $H_{\rm eff}$ it is easy to show that a hopping,
displacing an elementary triplet by a finite distance, starts in sixth
order in x, while in the 2-triplet sector $correlated$ hopping
starts in second order (see~\cite{knett00b} and Refs. therein),
 explaining the rather flat 1-triplet dispersion in contrast to the much
stronger pronounced 2-triplet dispersion as found by Kageyama et
al. by INS measurements~\cite{kagey00}. By means of degenerate
perturbation theory Momoi and Totsuka~\cite{momoi00b} also derive an effective
Hamiltonian for the Shastry-Sutherland model. Their third order result
also shows the significance of correlated hopping in the
multi-triplet dynamics of the model.\\
Let $|{\mathbf r}\rangle = |r_1,r_2\rangle$ denote the state of the 
system with one triplet on the dimer at ${\mathbf r}$ and
singlets on all other sites. The amplitude $t_{{\mathbf r'}- {\mathbf
    r}}^{o({\mathbf r})}$ for a 
triplet-hopping from site ${\mathbf r}$  to site ${\mathbf r'}$ is then 
given by 
\begin{equation}
\label{hopp_def}
t_{{\mathbf r'}-{\mathbf r}}^{o({\mathbf r})}= \langle{\mathbf
  r'}|{\cal H}_{\rm eff}|{\mathbf r}\rangle\ , 
\end{equation}
where the upper index $o({\mathbf r}) \in \{v,h\}$ allows to
distinguish whether the hopping started on a vertically oriented ($v$)
or a horizontally oriented dimer ($h$). Further we choose to split the
hopping amplitudes into a net part $\bar{t}_{\mathbf s}$ and a
deviation part $dt_{\mathbf s}$ (${\mathbf s}={\mathbf r'}-{\mathbf r}$)
\begin{equation}
  \label{split}
  t^{o({\mathbf r})}_{\mathbf s}=\bar{t}_{\mathbf s}+e^{iQ{\mathbf
  r}} dt_{\mathbf s}\ ,
\end{equation}
with $Q=(\pi,\pi)$.\\
Since $H_{\rm eff}$ conserves the number of triplets one has
\begin{equation}
  \label{H-eff_k}
  {H}_{\rm eff}|{\mathbf r}\rangle = \sum_{\mathbf r'}t^{o({\mathbf
  r})}_{{\mathbf r'}}|{\mathbf r}+{\mathbf r'}\rangle\ .  
\end{equation}
We introduce the Fourier transformed states
\begin{equation}
  \label{k_zust}
  |\sigma,{\mathbf k}\rangle=\frac{1}{{\sqrt N}}\sum_{\mathbf r}|{\mathbf r}\rangle
  e^{i({\mathbf k}+\sigma Q) {\mathbf r}}
\end{equation}
with the total  number of dimers $N$, the new quantum number $\sigma \in
\{0,1\}$ reflecting the sub lattice structure and ${\mathbf k}$ a
vector in the magnetic Brillouin zone (MBZ). ${H}_{\rm
  eff}$ acts as a $2\times 2$ matrix on the states $|\sigma,{\mathbf
  k}\rangle$ and $|1-\sigma,{\mathbf k}\rangle$. Its diagonalization yields
\begin{equation}
  \label{disp}
  \omega_{1/2}({\mathbf
  k})=\underbrace{\frac{a_0+a_1}{2}}_{\omega_0({\mathbf k})}\pm
  \frac{1}{2}\sqrt{(a_0-a_1)^2+4b^2}\ .
\end{equation}
Here we have defined
\begin{eqnarray}
  \nonumber
  a_{\sigma}&=& \left[\bar{t}_{\mathbf 0}+2\sum_{{\mathbf r} > 0}\bar{t}_{\mathbf
  r}\cos(({\mathbf k}+\sigma Q){\mathbf r}) \right]\mbox{ and}\\\label{a_b}
  b&=&2\!\!\!\!\!\!\sum_{{\mathbf r} > 0 \atop r_1+r_2\mbox{ \scriptsize even}}\!\!\!\!\!\!dt_{\mathbf r}\cos({\mathbf k}{\mathbf r})\ ,
\end{eqnarray}
with ${\mathbf r}>0$ if $r_1>0,$ or $r_1=0$ but $r_2>0$.
Thus the 1-triplet dispersion splits into two branches. We want to
point out, however, that at ${\mathbf k}={\mathbf 0}$ and on the
borders of the MBZ the two
branches fall onto each other leading to a 
2-fold degenerate dispersion at these points. This can be derived by
showing that the square root in Eq.~(\ref{disp}) vanishes at these
points~\cite{knett00c}. 
An analogous degeneracy is noticed 
in the 2-triplet sector. In Ref.~\cite{knett00b,knett00c} 
we give a detailed analysis of the symmetries of the model (2d space
group p4mm with underlying point group 4mm) and show that the
degeneracies are due to glide line symmetries.\\  
We calculated the amplitudes $\bar{t}_{\mathbf r}$ and $dt_{\mathbf
  r}$ (and therefore the dispersion) as exact polynomials in $J_1$, $x$ and
$y$ up to and including 15$^{\rm th}$ order.\\
Expanding the square root in Eq.~(\ref{disp}) about the limit of
vanishing $x$ and $y$ produces terms $\propto x^{\alpha}y^{\beta}$ with
$\alpha+\beta \ge 10$. Hence the energy splitting starts in
10$^{\rm th}$ order only. It is negligible for all reasonable values
of $x$ and $y$.\\
By substituting $y=0$ in $\omega_0({\mathbf k})$ we verify the
decimal numbers Weihong et al. \cite{weiho98b} obtained previously.\\
For three different sets of $J_1$, $x$ and $y$ the 1-triplet
dispersion is plotted in the inset of Fig.~\ref{merge1}. ESR~\cite{nojir99},
FIR~\cite{room99} and INS~\cite{kagey00} 
data suggest a value of $\omega({\mathbf 0})=2.98$meV. At finite ${\mathbf k}$
we have to rely on the INS measurement, which have rather large
errors. We get a very good agreement. But it is not
possible to fix the model parameters unambiguously from information in
the 1-triplet sector alone. Thus we continue our analysis in the
2-triplet sector.\\
The dynamics of two triplets  at large distances is governed by  1-triplet
hopping. At smaller distances a  2-particle
interaction occurs additionally given by $W_{h;{\bf d};{\bf r},{\bf d'}}$ 
($W_{v;{\bf d};{\bf r},{\bf d'}}$)
 starting with one triplet on a horizontal (vertical) dimer and another at
distance ${\bf d}$. The action of $H_{\rm eff}$ is to shift the triplets
to ${\bf r}$ and to ${\bf r}+{\bf d'}$. Nothing else is possible
due to  triplet number conservation. 
Since the total spin is conserved
($S\in\{0,1,2\}$) the distances are restricted to 
${\bf d}, {\bf d'} > {\bf 0}$, 
 because the exchange parity is fixed.\\
The coefficients $W$ for $S\in\{0,1,2\}$ and the 1-triplet hopping yield the complete 
2-particle dynamics. We compute $W$ up to $x^{12}$, the coefficients for 
the lowest-lying  states even  up  to $x^{14}$.\\
Analogously to the 1-triplet problem we use the following basis for
the 2-triplet states 
\begin{equation}
\label{basis}
|{\bf k},{\bf d},\sigma\rangle:=N^{-1/2}
\sum_{\bf r} 
e^{\left(i({\bf k}+\sigma{\bf Q})({\bf r}+{\bf d}/2)\right)}
 |{\bf r},{\bf r}+{\bf d}\rangle\ ,
\end{equation}
where ${\bf k}$ is the conserved total momentum in the MBZ applying
due to the two sub lattices; $|{\bf r},{\bf r}+{\bf d}\rangle $ denotes
the state with triplets at ${\bf r}$ and at ${\bf r}+{\bf d}$ (${\bf d}>0$). 
Since 1-triplet hopping is of higher order than the interaction
 an analytic expansion for the energies of 
the bound states is possible. At finite order in $x$  only
configurations contribute where the two triplets are
not too far away from each other.
 Of course, higher orders imply larger, but still finite distances.
 In particular, the energies of the four states which evolve from 
neighboring triplets can be computed very well since their
leading interaction is linear. 
Investigating the
matrix elements shows that it is sufficient to study the distances
${\bf d}\in\{(0,1),(1,0),(1,\pm1)\}$ 
for order 5. To $x^{14}$ only 
${\bf d}\in\{(1,\pm2),(2,\pm1),(0,2),(2,0),(2,\pm2)\}$
must be added. So, for given total momentum only a finite 
$8\times 8$ or $24\times 24$ matrix has to be analyzed. Furthermore,
the  elements connecting shorter distances to longer distances
and the elements among longer distances do not need to be known to 
very high order.\\
We have analyzed the dispersions in $x^5$
of the four  states bound linearly in $x$ in the MBZ. 
Fukumoto's results are mostly confirmed \cite{fukum00}. The
dispersion of bound states  starts only in  $x^3$
(contrary to $x^4$ claimed in Ref.~\cite{lemme00a}).
At particular points of high point group symmetry 
($(0,0)$,$(0,\pi)$)
 the  Hamiltonian splits into six blocks corresponding
to different representations of the square  point group 4mm. At these points
the analysis up to $x^{14}$ is carried out.
 The symmetries are classified according to the
irreducible representations (four 1D, one 2D) 
of the point group 4mm
$\Gamma_1 (1), \Gamma_2 (x^2-y^2), \Gamma_3 (xy), \Gamma_4 (xy(x^2-y^2)), 
\Gamma_5 (x, y)$ where simple polynomials are given in brackets to show
 the transformation behavior.
The extrapolated energies are depicted in Figs.~\ref{merge1} ($S=0$)
 and \ref{merge2} ($S=1$) as functions of $x$. The double degeneracy
 for ${\bf k}=(0,\pi)$ results from the same symmetry reasons as
 described for the 1-magnon dispersion. The dashed-dotted curve at $(0,\pi)$ has to
be compared to the solid and the long-dashed curve to assess the 
dispersion of these two modes from ${\bf 0}$ to $(0,\pi)$.
While for $S=0$ this dispersion always has the expected
behavior with $\omega({\bf 0}) < \omega((0,\pi))$ the energies for $S=1$
are reversed  for small values of $x$ (cf. \cite{fukum00}). 
Only above $x\approx 0.55$ the relation 
$\omega({\bf 0}) < \omega((0,\pi))$ holds for $S=1$.
For $S=0$, the lowest mode vanishes at the same
$x$ as does the elementary triplet gap $\Delta$. So no additional instability
occurs for $S=0$. This provides evidence against the competing 
singlet phase as presumed in Ref.~\cite{koga00}. There is, however, a salient instability for $S=1$ (Fig.~\ref{merge2}) at
$x=0.63$. This comes as a surprise since one expects in antiferromagnets
binding effects to be largest for $S=0$ (discussed in more detail in
Ref.~\cite{knett00b}).\\
Assuming $\Delta=2.98$meV as above and $\omega |_{S=1}=4.7$meV from
ESR~\cite{nojir99}, FIR~\cite{room99} and INS~\cite{kagey00} we can
determine the model parameters for 
SrCu$_2$(BO$_3$)$_2$ precisely ($y=0$): $x=0.603(3)$ and
$J_1=6.16(10)$meV.\\
The inset in Fig.~\ref{merge1} shows that the
1-magnon dispersion agrees very nicely with experiment for these
values. Further, the energy of the $\Gamma_3$ singlet matches the
30cm$^{-1}$ peak in Raman scattering~\cite{lemme00a} perfectly. The
$\Gamma_4$ singlet at 25cm$^{-1}$ is forbidden by symmetry, since the
Raman operator is $\Gamma_3$ at T=0~\cite{knett00b}. Calculating the
next $\Gamma_3, S=0$ bound state 
(not shown) yields 45cm$^{-1}$ in good agreement with the
experimental 46cm$^{-1}$ line, too.\\
We conclude that the 2D model (Fig.~\ref{model}) explains the low-lying
excitations of SrCu$_2$(BO$_3$)$_2$ perfectly. 
Thermodynamic quantities like the susceptibility
$\chi(T)$ require the inclusion of the interplane coupling $J_\perp$
which is fully frustrated not changing the dimer spins
\cite{miyah00b}. The 3D $\chi_{\rm 3D}$ is computed
from $\chi_{\rm 2D}$ on the  mean-field level as
$\chi_{\rm 3D}^{-1}=\chi_{\rm 2D}^{-1}+4J_\perp$ (inset in
Fig.~\ref{merge2}). Fits to the experiment~\cite{kagey99} give $J_\perp=1.3(2)$meV 
leading to a curve agreeing without flaw above 40K. Our value for
$J_\perp$ is significantly 
higher than the one in Ref.~\cite{miyah00b} due to different values of
$x$ and $J_1$.\\
Summarizing, we presented the first
quantitative description of 2-particle bound states in 2D.\\
An unexpected instability for the $S=1$ 2-triplet bound state
  is found at $x\approx0.63$ indicating a transition to
a triplet condensate probably related to the helical phase found previously
\cite{albre96,mulle00}. We conjecture that this transition is first
order  occuring at lower $x$ than assumed so far.\\
The symmetries of experimentally relevant states were determined.
The reliability of the high order results allows to fix the 
experimental coupling constants very precisely 
($J_1=6.16(10)$mev, $J_2/J_1=0.603(3)$, $J_\perp=1.3(2)$meV). 
Thereby, various experiments (ESR, FIR, INS, Raman, $\chi(T)$) 
are explained consistently.\\
The work is supported by
the DFG in SFB 341 and in SP 1073.

\end{document}